\documentclass[sigconf,natbib=true]{acmart}
\usepackage{amsmath,amsfonts}
\usepackage{adjustbox}

\usepackage{algorithm}
\usepackage[noend]{algpseudocode}
\usepackage{graphicx}
\usepackage{textcomp}
\usepackage{xcolor}
\usepackage{caption}
\usepackage{subcaption}
\usepackage{tabularx}
\usepackage{multirow}
\usepackage{makecell}
\usepackage{diagbox}
\usepackage{booktabs}
\usepackage{color, colortbl}
\usepackage{courier}
\usepackage{pifont}
\usepackage{enumitem}
\definecolor{Gray}{gray}{0.9}
\algrenewcommand\algorithmicrequire{\textbf{Input:}}
\algrenewcommand\algorithmicensure{\textbf{Output:}}

\AtBeginDocument{
  \providecommand\BibTeX{{
    \normalfont B\kern-0.5em{\scshape i\kern-0.25em b}\kern-0.8em\TeX}}}

\setcopyright{acmcopyright}
\copyrightyear{2024}
\acmYear{2024}

\author{Yoonhyuk Choi, K. Sel\c{c}uk Candan, Huan Liu }
\orcid{0000-0003-4359-5596}
\affiliation{
  \institution{
  Arizona State University}
  \city{Tempe}
  \state{Arizona, USA}
}
\email{jim1995,candan,hliu@asu.edu}

\author{Reepal Shah, John Sabo}
\orcid{0000-0003-4359-5596}
\affiliation{
  \institution{
  Tulane University}
  \city{New Orleans}
  \state{Louisiana, USA}
}
\email{rshah3,jsabo1@tulane.edu}

\renewcommand{\shortauthors}{Yoonhyuk Choi et al.}

\newtheorem{manualtheoreminner}{\textbf{Theorem}}
\newenvironment{manualtheorem}[1]{%
  \renewcommand\themanualtheoreminner{#1}%
  \manualtheoreminner
}{\endmanualtheoreminner}

\acmConference[CIKM '24] {Proceedings of the 33rd ACM International Conference on Information and Knowledge Management}{October 21--25, 2024}{Boise, Idaho, USA}
\acmBooktitle{Proceedings of the 33rd ACM International Conference on Information and Knowledge Management (CIKM '24), October 21--25, 2024, Boise, Idaho, USA}
\acmPrice{15.00}
\acmISBN{978-1-4503-9236-5/22/10}
\acmDOI{10.1145/3511808.3557324}

\begin{document}

\title{Prioritizing Potential Wetland Areas via Region-to-Region Knowledge Transfer and Adaptive Propagation}

\begin{abstract} 
Wetlands are important to communities, offering benefits ranging from water purification, and flood protection to recreation and tourism. Therefore, identifying and prioritizing potential wetland areas is a critical decision problem. While data-driven solutions are feasible, this is complicated by significant data sparsity due to the low proportion of wetlands (3-6\%) in many areas of interest in the southwestern US. This makes it hard to develop data-driven models that can help guide the identification of additional wetland areas. To solve this limitation, we propose two strategies: (1) The first of these is knowledge transfer from regions with rich wetlands (such as the Eastern US) to sparser regions (such as the Southwestern area with few wetlands). Recognizing that these regions are likely to be very different from each other in terms of soil characteristics, population distribution, and land use, we propose a domain disentanglement strategy that identifies and transfers only the applicable aspects of the learned model.    
(2) We complement this with a spatial data enrichment strategy that relies on an adaptive propagation mechanism. This mechanism differentiates between node pairs that have positive and negative impacts on each other for Graph Neural Networks (GNNs).
To summarize, given two spatial cells belonging to different regions, we identify domain-specific and domain-shareable features, and, for each region, we rely on adaptive propagation to enrich features with the features of surrounding cells.
We conduct rigorous experiments to substantiate our proposed method's effectiveness, robustness, and scalability compared to state-of-the-art baselines. Additionally, an ablation study demonstrates that each module is essential in prioritizing potential wetlands, which justifies our assumption.
\end{abstract}

\ccsdesc[500]{Information systems~Information systems applications~Spatial-temporal systems~Geographic information systems}

\keywords{Recommender system, Wetland identification, Transfer learning, Domain disentanglement, Adaptive propagation, Data sparsity}

\maketitle

\begin{figure}[h]
  \vspace{5mm}
  \centering
  \includegraphics[width=0.43\textwidth]{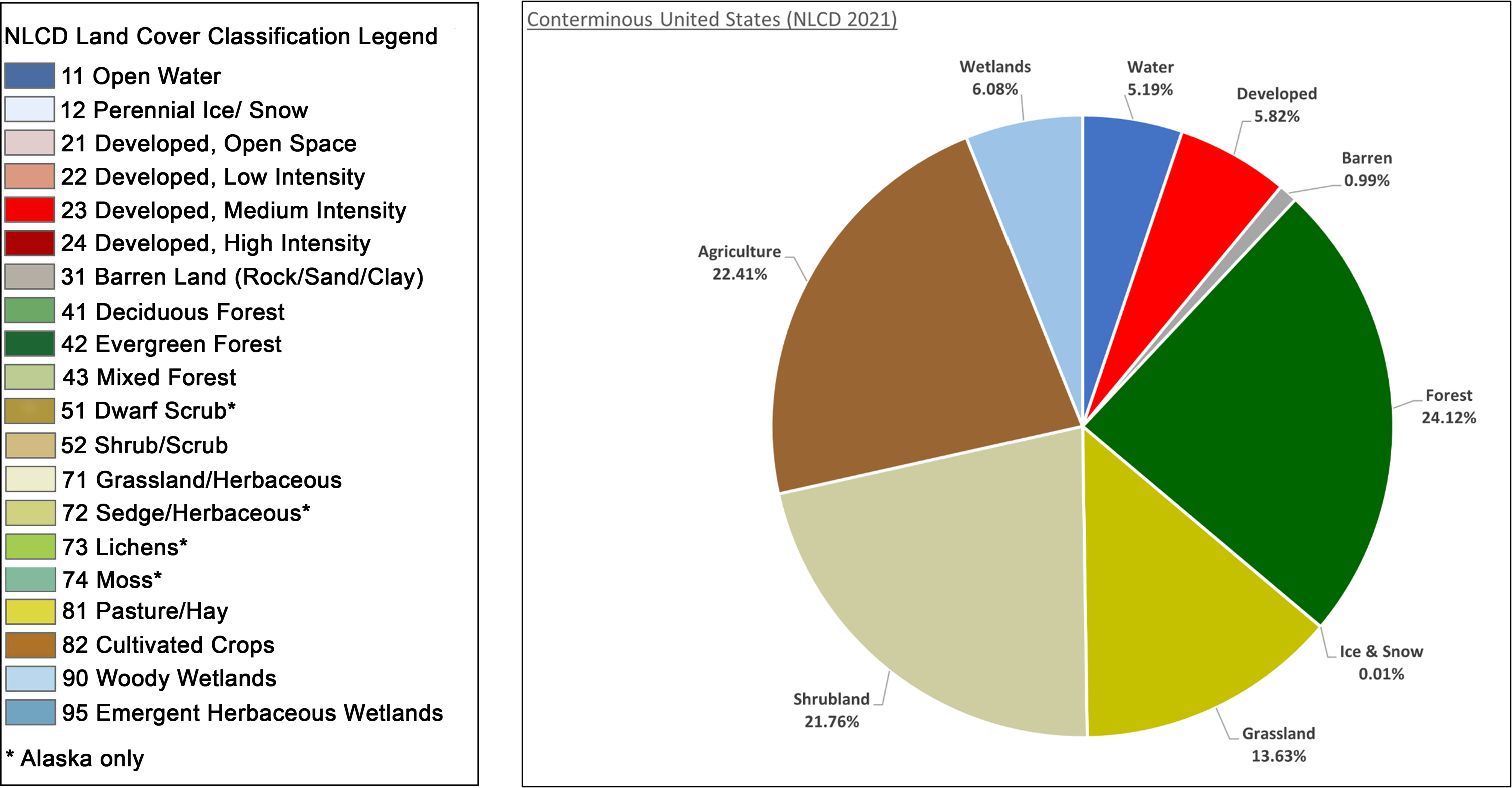}
  \caption{Natural Land Cover Dataset (NLCD \cite{dewitz2021national} provides 20 categories of land use, including wetlands (types 90 and 95). The data reveals that only 6\% of the land in the US is wetland}
  \label{fig:nlcd}
\end{figure}

\section{Introduction}
Wetlands, characterized by the presence of water at or near the surface of the soil for most of the year, play several crucial roles within an ecosystem. Specifically, they provide habitat for various plant and animal species, improving water quality by filtering pollutants and storing carbon that mitigates climate change \cite{lyon1993wetland}. Furthermore, wetlands are also important to communities for water purification and flood protection \cite{hammer2020wetlands}. Despite their value to the ecosystem, wetlands face challenges from competition with agriculture and urban development, as well as from climate change. These highlight the need for both the conservation of existing wetlands and the identification of potential new wetland areas\footnote{This research has been funded by 
NSF$\#$2311716 and by the US Army Corps of Engineers Engineering With Nature Initiative through Cooperative Ecosystem Studies Unit Agreement $\#$W912HZ-21-2-0040}. 

\subsection{Challenges and Solutions}

A major challenge in developing new wetlands, particularly in areas where they are essential for water conservation, is the arid climate and local soil features that make water retention difficult. As shown in Figure \ref{fig:nlcd}, wetlands only account for 6\% of land cover in the US and the share of wetlands gets even lower if in dryer regions. 
The sparsity of wetlands makes it hard to develop data-driven models that can help guide the identification of new wetland areas.
In particular, given the large imbalance between the wetland and non-wetland regions, 
naively developed models can suffer from overfitting and bias issues. This issue is further complicated considering that regions with few wetlands are not necessarily suitable for wetland development, making it particularly challenging to develop data-driven models with sparse positive samples.

\begin{figure}[t]
  \vspace{5mm}
  \centering
  \includegraphics[width=0.3\textwidth]{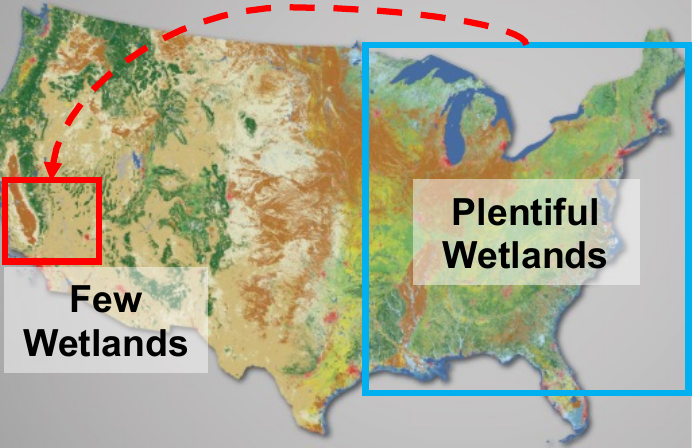}
  \caption{Knowledge transfer from wetland-rich areas of the US to regions where wetlands are inherently sparse}
  \label{fig:kt}
\end{figure}

In this paper, we argue that this {\em data sparsity} challenge can be addressed by relying on two complementary techniques:
\begin{itemize}[leftmargin=*]
\item {\em Region-to-region knowledge transfer:} As illustrated in Figure \ref{fig:kt}, we can transfer knowledge from regions with rich wetlands (such as the Eastern US) to regions where wetlands are sparse (such as the Southwestern US with few wetlands). However, this solution suffers from {\em {\bf global context incompatibility} challenge -- } since the source and target regions are likely to differ significantly in soil characteristics, population distribution, and land use. Moreover, the expected outcomes (water storage vs. flood prevention) from the wetlands make it difficult to transfer knowledge from the source context to the target.  
\item {\em Adaptive knowledge transfer within local regions:} We complement this with a spatial data enrichment strategy that transfers useful information among the spatial cells, with different characteristics, in the same vicinity. This solution also encounters a form of {\em {\bf local context incompatibility,}} where information from nearby cells should be considered differently: while some cell pairs {\em positively} affect each other, others have a {\em negative} impact. Hence, it's crucial to identify and consider this aspect when implementing knowledge transfer between local regions.

\end{itemize}

We tackle the first of these challenges, namely global context incompatibility,  through {\em domain disentanglement,} where we separate domain-specific information from information shareable across regions.
More specifically, we leverage a domain discriminator-based approach which disentangles domain-specific and domain-shareable features more effectively than entropy-based approaches. 

We tackle the second, local context incompatibility, challenge, by proposing an  {\em adaptive propagation}  technique that adjusts the weight of message-passing in a way that differentiates between node pairs that have positive and negative impacts. 
Graph Neural Networks (GNNs), commonly used for local knowledge transfer, generally assume homophily, wherein connected (or adjacent) nodes are likely to share the same label. 
However, this assumption is especially problematic where we have very few wetlands in a given region, and applying homophilic GNNs could significantly degrade performance by overwhelming information provided by these few wetlands with information from plentiful non-wetland regions. 
Therefore, we propose an adaptive propagation technique that appropriately adjusts information passed from neighboring nodes or uses information from further away nodes to prevent bias in local knowledge transfer.

\subsection{Contributions of this Paper}

In summary, we propose a novel method of identifying {\em \textbf{Po}tential wetlands via \textbf{T}ransfer learning and \textbf{A}daptive propagation (PoTA)}
to tackle the data sparsity challenge in wetland prioritization problem, which is equipped with two complementary knowledge transfer techniques: 
\begin{itemize}
\item We suggest a long-range, region-to-region knowledge transfer from locations rich in wetlands to sparser ones using domain disentanglement 
\item Coupled with adaptive feature propagation within individual regions, we employ an enhanced message-passing scheme between heterophilic neighbors.
\end{itemize}
We provide theoretical analyses of the effectiveness of the proposed components in solving data sparsity issues.
We further conduct experiments to test the empirical efficacy of our method considering six regions of the US, with sparse to moderate to dense wetland portfolios and varying soil and land use characteristics. Our empirical investigation includes ablation studies to explore the contributions of the various modules in solving this problem.

\section{Related Work}
\subsection{Wetland Identification}
Wetlands are known to occur due to reasons such as permanent inundation or soil saturation \cite{tiner1991concept}. Due to
the value of their resources, prioritizing potential wetlands has become an important topic \cite{lyon1993wetland}. Generally, the formation of wetlands is known to be closely related to soil characteristics \cite{faulkner1989field,megonigal1993wetland,mausbach1994classification,richardson2001wetland}. Therefore, various methods have been proposed for wetland identification, such as collecting geographic data based on sensing \cite{adam2010multispectral,mahdavi2018remote} or finding indicators \cite{tiner2016wetland} necessary for wetland formation. Based on these associations, the most recent approach presented is solving the problem using deep learning techniques on the physically-informed data \cite{o2020deep,greenhill2024machine}.

\subsection{Transfer Learning}
Knowledge transfer (a.k.a transfer learning) refers to using additional data (source domain) to solve data sparsity when training data is lacking in the target domain. Generally, this method utilizes the common contexts that exist in both domains, which is widely used in text classification \cite{do2005transfer}, digit recognition \cite{maitra2015cnn}, and recommender systems \cite{choi2022based}. The common context means some information like images, digits, ratings, and so on. In our case, the natural features like soil and drainage features from different regions could be used for transfer learning. Here, we focus on the cross-domain recommendation schemes \cite{man2017cross,wang2018cross, zhao2020catn,zhu2021cross}, which suggest capturing common knowledge from both domains. The foundational concept called domain adaptation \cite{yuan2019darec,bonab2021cross} has been proposed, which captures domain-shareable features (common knowledge) through adversarial training. However, several recent studies have revealed that the common features without specific guidance may not always be helpful to the target domain \cite{peng2019domain,li2019learning,nema2021disentangling}. This becomes more severe as the discrepancy between domains (e.g., different categories) increases, leading to the proposal of domain disentanglement techniques \cite{cai2019learning,cheng2020improving,cao2022disencdr,choi2024based} to address this issue.

\subsection{Adaptive Propagation}
Information propagation \cite{neumann2016propagation} is a widely used technique in graph theory. In detail, the mechanism of Graph Neural Networks (GNNs) is node embedding and message-passing (propagation) \cite{garcia2017learning}, which has the advantage of using the adjacent nodes for the prediction. However, message-passing algorithms may fail to perform well under heterophilic graphs \cite{pei2020geom}, where most edges connect two nodes with different labels. To conquer this problem, several studies suggested finding these connections \cite{huang2019signed,choi2022finding,zhao2023graph} by measuring the difference of nodes (e.g., attention) or by utilizing remote nodes with high similarity (non-local aggregation). In addition, \cite{zhu2020beyond} proposes ego-neighbor separation for message-passing, \cite{zhu2021graph} generates a compatibility matrix, \cite{li2022finding} further utilizes neighbors (non-local propagation), and \cite{ijcai2022p310} figures out the path-level pattern. As another branch, the mechanism of adaptive propagation \cite{wang2022powerfulb}, and choosing appropriate architectures \cite{zheng2023auto} have been recently proposed. The core concept of adaptive propagation is that they determine the sign of edges \cite{chien2020adaptive,bo2021beyond,fang2022polarized,guo2022clenshaw} before applying message-passing scheme, which can be either positive or negative.

\section{Preliminaries}
In this section, we present the preliminaries, including useful notations that will be employed throughout this paper. 
We first introduce the natural features available for wetland prioritization. Then, we formulate the wetland prioritization problem.

\subsection{Land Cover and Surface Features} \label{data_source}
In this paper, we primarily rely on the Natural Land Cover Dataset (\textbf{NLCD 2021}\footnote{https://www.indianamap.org/maps/INMap::nlcd-land-cover-2021/about}) visualized in Figure \ref{fig:nlcd}.
The data set includes the land cover type (20 categories, at $30m\times 30m$ resolution)
from 2001 to 2021. 
This data set is complemented with soil type information, from 
\textbf{SSURGO}\footnote{https://www.nrcs.usda.gov/resources/data-and-reports/gridded-soil-survey-geographic-gssurgo-database} data set, and drainage information from {\em Height Above the Nearest Drainage} \textbf{(HAND)} data set\footnote{https://registry.opendata.aws/glo-30-hand/}. The SSURGO feature consists of soil characteristics, slope gradient, water table depth, available water storage, and so on.
Note that these two data sets have different resolutions from the NLCD data set -- for instance, the SSURGO data set has a finer $10m\times 10m$ resolution. Therefore, we integrated the three data sets by downscaling their resolutions as appropriate. The final aligned data sets have $30m\times 30m$ resolution.

\begin{table}[t]\caption{Notations}
\vspace{-1.5em}
\small
\begin{center}
     \begin{tabular}{@{}l|l@{}}
\toprule
\multicolumn{1}{c|}{\textbf{Symbol}} & \textbf{Explanation} \\
\toprule
\multicolumn{1}{c|}{$\mathcal{S},\mathcal{T}$} & Source and target domains\\
\multicolumn{1}{c|}{$\mathcal{G}$} & Adjacency graph of cells\\
\multicolumn{1}{c|}{$c$} & Specific cell\\
\multicolumn{1}{c|}{$w_e=c, \widehat{w}_c$} & True and predicted wetland label\\
\multicolumn{1}{c|}{$s_c$} & Multi-dimensional soil features \\
\multicolumn{1}{c|}{$h_c$} & Height above nearest drainage (binary)\\
\multicolumn{1}{c|}{$l_c$} & Extracted latent feature of $s_c,h_c$ \\
\multicolumn{1}{c|}{$d_c, \widehat{d}_c$} & True and predicted domain label of $f_c$\\
\multicolumn{1}{c|}{$N_s,N_t$} & Mini-batch training samples for $s$ and $t$ \\
\multicolumn{1}{c|}{$\oplus$} & Concatenation operator \\
\multicolumn{1}{c|}{$\mathcal{L}$} & Loss function \\
\bottomrule
    \end{tabular}
\end{center}
    \label{tab:notations}
\end{table}


\subsection{Problem Formulation} \label{prob_for}
The wetland prioritization task can be formulated as follows. As mentioned above, for each target region ($R_\tau$), we have land cover (NLCD), soil (SSURGO), and drainage (HAND) data. Specifically, for each cell $c$ in the range, we have a wetland label $w_c \in \{0, 1\}$ (wetland if $w_c=1$), multi-dimensional soil feature vector $s_c$, and drainage flag value $h_c \in \{0, 1\}$. 
Coupled with these, we further assume an adjacency matrix $\mathcal{G}_\tau,$ which describes the neighborhood relationship among the cells in the region. Given these, we define the {\em wetland classification} problem as follows:
\begin{definition}[Wetland Classification]
Given the set of tuples  $D_\tau = \{ (c, \mathcal{G}_c, s_c, h_c)\;|\; c\in R_\tau\}$, the {\em wetland classification} problem aims to find a mapping function $f_\tau(c, \mathcal{G}_c, s_c, h_c) \rightarrow w_c$.
\end{definition}

The careful reader would note that the wetland classification problem, in and of itself, will not enable wetland identification and prioritization tasks as the only thing a perfect model with $100\%$ accuracy would provide is an explanation of the current wetlands in the given region, rather than recommending new wetlands. Therefore, we seek models with {\em high recalls} (i.e., accurately explaining existing wetlands in the region), while not necessarily having perfect precision. For instance, a cell $c$ in the target region $R_\tau$ where the model predicts a wetland, despite the absence of one currently, may serve as a candidate for new wetland development.

As discussed before, our goal is to improve the wetland identification and prioritization task for target regions with sparse wetlands by relying on information obtained from source regions with denser wetlands. Therefore, we present a slightly modified version of the problem as follows:

\begin{definition}[Wetland Classification with Knowledge Transfer]
Given a source region, $R_{s}$ and a target region $R_{\tau}$ and the associated sets of tuples $D_{s}$ and $D_{\tau}$, the {\em wetland classification with knowledge transfer} problem aims to find a mapping function $f_{s,\tau}(c, \mathcal{G}_c, s_c, h_c) \rightarrow w_c$, where $c \in R_\tau$.
\end{definition}

As before we seek models with {\em high recalls} for the target region. High precision is preferred, with the understanding that a cell $c \in R_\tau$ for which $w_c = 0$ and $f_{s,\tau}(c, \mathcal{G}_c, s_c, h_c) = 1$ may be a candidate for new wetland development.  
Note that, while the classification problem has been defined considering those cells in the target region (i.e., $c \in R_\tau$), as a potential by-product of the process, the learned mapping function may be used to classify also the cells in the source region. The notations used throughout this paper are summarized in Table \ref{tab:notations}.

\begin{figure}[t]
  \includegraphics[width=\columnwidth]{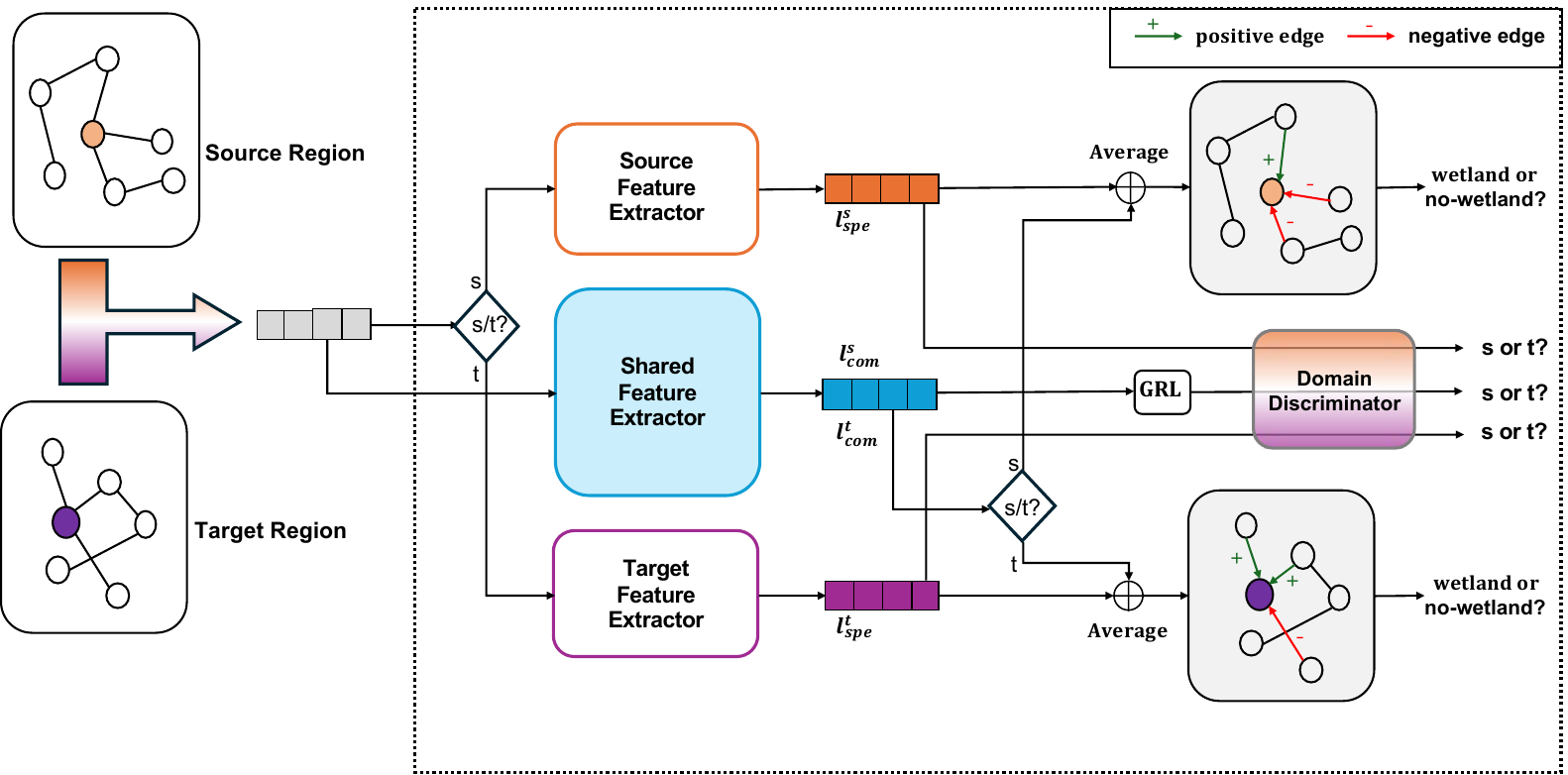}
  \caption{Overview of the proposed  {\em \textbf{Po}tential wetlands via \textbf{T}ransfer learning and \textbf{A}daptive propagation (PoTA)} }
  \label{fig:model}
\end{figure}

\section{Methodology}
In this section, we describe our {\em \textbf{Po}tential wetlands via \textbf{T}ransfer learning and \textbf{A}daptive propagation (PoTA)} algorithm, designed to tackle the data sparsity challenge in wetland prioritization.
Figure \ref{fig:model} provides an overview of the overall architecture, where we introduce the key components of the model below: 
\begin{itemize}[leftmargin=*]
    \item \textbf{(Feature Processing -- Section~\ref{sec:feature})} 
    Input features (for both source and target regions) are vectorized through discretization or normalization based on their types. 
    \item \textbf{(Transfer Learning -- Section~\ref{sec:kt})} Three latent feature extractors (LFEs) are applied to these features to capture domain-aware knowledge. Specifically, a domain discriminator aids the extraction of two sets of (source and target) domain-specific latent features along with one set of shared latent features. The shared latent features serve as the bridge that enables knowledge transfer from the (wetland-rich) source region to the target region.
    \item \textbf{(Adaptive Propagation -- Section~\ref{sec:ap})} In the next step, we apply adaptive propagation on top of the extracted latent features. 
    In particular, the proposed algorithm learns whether information propagation between the two cells should be positive or whether the information impact of one cell on another should be negative.
\end{itemize}
Finally, the combined latent features train wetland classifiers for the source and target regions.

\subsection{\textbf{Feature Processing and Extraction}}\label{sec:feature}

\subsubsection{Soil Data} This dataset includes discrete classes like flood frequency and elements with continuous values, like available water storage. We encode these features as follows:
\begin{itemize}
    \item \textbf{Discrete features} are one-hot encoded. 
 Note that, some discrete features, like flood frequency, are ordinal rather than categorical. In this implementation, we ignore the distinction between these two types of discrete values and use one-hot encoding for both. 
    \item \textbf{Continuous features,} such as water storage, are normalized to the range $[0,1]$ by dividing to the maximum value among all the cells in the source and target regions. 
\end{itemize}

\subsubsection{HAND Data} Based on the domain expert feedback, the HAND dataset is binarized using a $2m$ threshold. Specifically, for any cell where the height above the nearest drainage is greater than or equal to $2m$, the value of $h_c$ is set to $1$; otherwise, $h_c$ is set to 0.

\subsection{Transfer Learning}\label{sec:kt}
In this section, we introduce the transfer learning techniques we developed to effectively bring information learned from the wetland-rich source region to the target region with few wetlands. 
In particular, the encoded input feature vectors for both source and target regions are fed into the two layers of neural networks, referred to as latent feature extractors (LFEs). As shown in Figure \ref{fig:model}, these three LFEs generate four outputs; source-domain specific latent features of the source ($l^s_{spe}$), 
target-domain specific latent features of the target ($l^t_{spe}$),
shared-domain (or common-domain) latent features of the source ($l^s_{com}$),
and shared-domain latent features of the target ($l^t_{com}$). In this subsection, we present detailed strategies to guide this process for effectively extracting domain-specific and shared-domain latent features.

Domain Adaptation \cite{ganin2016domain} aims to reduce the differences in latent features extracted from two domains. 
While such an approach may help find domain shareable features, we also need to identify the unique characteristics of regions for wetlands (e.g., Arizona is extremely dry due to its desert climate) \cite{chen2019transferability}. Many approaches have been proposed recently \cite{gretton2005measuring,li2019learning,nema2021disentangling,choi2024based}, but in PoTA, we develop a domain disentanglement strategy, as in \cite{choi2022based}, to capture domain-specific knowledge along with shareable information.
To achieve this, a Gradient Reversal Layer (GRL), which applies a negative multiplier on the back-propagated weights during training,  is used before the extracted latent features are fed into a domain discriminator to increase the entropy between the learned latent features. 
The key is that by applying GRL only to shared-domain features, the domain discriminator is trained to distinguish between the specific characteristics of the two domains effectively.
PoTA leverages two layers of fully connected neural networks as the domain discriminator. As can be seen in  Figure \ref{fig:model}, for each cell in the source or target region, two features, $l_{com}^*$ and   $l_{spe}^*$ (where $* \in \{s,t\}$) are passed separately as inputs to the domain discriminator:
\begin{equation} 
\label{domain_label}
\widehat{d}_{com}^*=F_{disc}(l_{com}^*),\, \,\widehat{d}_{spe}^*=F_{disc}(l_{spe}^*).
\end{equation}
Here, $\widehat{d}^*$ stands for the predicted domain probability of the input latent feature.  
The loss, then, is calculated through binary cross-entropy as  follows: 
\begin{equation}
\begin{gathered}
\mathcal{L}_{com}^*= -{1 \over N_*}\sum_{s=1}^{N_*} log(1-\widehat{d}^{*}_{com}),
\\
\mathcal{L}_{spe}^*= -{1 \over N_*}\sum_{s=1}^{N_*} log(1-\widehat{d}^*_{spe}),
\end{gathered}
\label{d_inv_loss}
\end{equation}
where $N_s$ and $N_t$ are the training batches for the source and target regions respectively (we omit the true label in the above equation that is binary for the source and target domain, $d \in \{0, 1\}$). 
Given this, we define the overall domain loss as,
\begin{equation}
\begin{gathered}
\mathcal{L}_{dom} = \rho(\mathcal{L}_{com}^s +
\mathcal{L}_{spe}^s) + (1-\rho)(\mathcal{L}_{com}^t +  \mathcal{L}_{spe}^t),
\end{gathered}
\label{da_loss}
\end{equation}
Here, the ratio $\rho={N_s \over N_s+N_t}$ of the batch sizes serve as the weight for the training of the discriminator concerning the importance of the source domain \cite{ben2010theory}. During training, the domain discriminator is updated using Equation \ref{da_loss}. 

\subsection{Adaptive Propagation}\label{sec:ap}

The knowledge transfer between local regions is facilitated by a graph neural network (GNN). Let $\mathcal{G}^*=(\mathcal{V}^*,\mathcal{E}^*, X^*)$ be a graph with $\vert\mathcal{V}^*\vert=n$ nodes, $\vert\mathcal{E}^*\vert=m$ edges, and feature matrix $X^*$. As before, when considering the source region, $*  = s$, and when considering the target region, we have $* = t$. 

The label matrix corresponding to the given region is denoted as $Y^* \in \mathbb{R}^{n \times C}$, where $C (= 2)$ stands for the number of classes. $A^* \in \{0, 1\}^{n \times n}$ denotes the adjacency matrix for the undirected graph $\mathcal{G}^*$, where the degree of node $i$ is represented as $\delta_{i}=\sum^n_{j=1}{A^*_{ij}}$. The feature matrix, $X^* \in \mathbb{R}^{n \times h}$,  where $h$ is the number of dimensions of the latent space. 
Given the above, the representation of $\mathcal{V}^*$ is updated through message-passing between neighboring nodes.

GNNs have the advantage of addressing data sparsity by incorporating the characteristics of neighboring nodes into their predictions. However, if neighboring cells have very different characteristics, this can lead to performance degradation as the propagated information may serve as noise, rather than enrichment. Therefore, here we propose an adaptive propagation technique to address this issue. 
This is achieved by varying the edge weights considering the features of the two neighboring cells corresponding to the edge. 
We first average the domain-specific and shared-domain features for all nodes in the given region (i.e., $v_i \in \mathcal{V}^*$):
\begin{equation}
    l_i^*={l^*_{spe,i} + l^*_{com,i} \over 2}, 
\end{equation}

Next, given an edge between $i \leftrightarrow j$, we determine the corresponding edge weight through an attention mechanism:
\begin{equation}
    w^*_{ij}=tanh\left((l^*_i + l^*_j)\Vec{a}_*\right),
\end{equation}
where $\Vec{a}_*$ is a learnable attention vector; $tanh$ helps ensure that edge weights satisfy the constraint $-1 \leq w^*_{ij} \leq 1$. Given these edge coefficients, following~\cite{bo2021beyond}, we update  node features as follows:
\begin{equation}
\label{message_passing}
    l^{*, (\gamma)}_i=l_i^{*, (0)}+\sum_{j \in \mathcal{N}_i}{w^*_{ij} \over \sqrt{\delta_i \delta_j}}l^{*, (\gamma-1)}_j,
\end{equation}
where, $\mathcal{N}_i$ denotes the set of neighbors of node $i$ with incoming edges into $i$. As defined earlier,  $\delta$ and $(\gamma)$  denote the degree of a node and hidden layer, respectively. 

\subsection{Classification}
Assuming $L$ layers of propagation, the final output of the adaptive propagation is an enriched latent representation $l_i^{*, (L)}$ for $v_i \in \mathcal{V}^*$. This is fed into a fully connected network, $W^* \in \mathcal{R}^{h \times C}$, with input dimensionality $h$, and output dimension of $C (=2)$, corresponding to "wetland" and "no-wetland", respectively.

Given the final output, $f^{*, (L)} \in \mathcal{R}^{N_* \times C}$ (where $N_s$ and $N_t$ are the training batches for the source and target regions respectively), the wetland probability is given by:
\[f^{*, (L)} = l^{*, (L)} W^*,\]
where we can define the prediction loss using the negative log-likelihood function as follows:
\begin{equation}
\label{pred_loss}
    \mathcal{L}^*_{pred}=-{1 \over N_*} \sum^{N_*}_{s=1}\sum^C_{k=1}y_{sk} \log (f^{*, (L)}_{sk}).
\end{equation}

\subsection{Optimization and Inference}

\textbf{(Optimization)} We optimize the model using domain loss, $\mathcal{L}_{dom}$ (Eq. \ref{da_loss}), and prediction loss, $\mathcal{L}_{pred}$ (Eq. \ref{pred_loss}) as below:
\begin{equation}
\label{overall_loss}
    \mathcal{L}=\mathcal{L}^s_{pred}+\mathcal{L}^t_{pred}+\lambda \mathcal{L}_{dom}.
\end{equation}
Here, $\lambda$ adjusts the weight of domain loss to help stable convergence. During optimization, we employ the Adam optimizer with early stopping based on the validation score. 

\textbf{(Inference)} After convergence, we forward the features of each cell in a target domain to the common and target LFEs, followed by the adaptive GNN for wetland prioritization.

\subsection{Theoretical Analysis} \label{theo_anal}
This section provides theoretical background on why domain disentanglement and signed propagation are necessary for this task.
\begin{manualtheorem}{4.1}[Domain disentanglement]\label{thm1}
Let us assume a domain identifier $\mathcal{D} \in \{\mathcal{S},\mathcal{T}\}$. Regardless of a specific domain $\mathcal{D}$, the mutual information (\textbf{I}) between the domain-common feature $l^*_{com}$ and domain-specific one $l^*_{spe}$ \cite{hwang2020variational} can be decomposed as below:
\begin{equation}
    I(l^*_{spe};l^*_{com})=-I(\mathcal{D};l^*_{spe},l^*_{com})+I(\mathcal{D};l^*_{spe})+I(\mathcal{D};l^*_{com})
\end{equation}
Then, we get $I(\mathcal{D};l^*_{spe},l^*_{com})=I(\mathcal{D};l^*_{spe})+I(\mathcal{D};l^*_{com})-I(l^*_{spe};l^*_{com})$ with a slight modification. Since $I(\mathcal{D};l^*_{com})$ is maximized by a domain discriminator and $I(l^*_{spe};l^*_{com})$ is minimized through the objective function (wetland identification), we can infer $I(\mathcal{D};l^*_{spe},l^*_{com}) \geq I(\mathcal{D};l^*_{spe})$ that is more informative for training.
\end{manualtheorem}

\begin{manualtheorem}{4.2}[Adaptive propagation]\label{thm2}
Adaptive propagation determines the weight and sign of edges based on features. While a positive signed edge has the effect of smoothing between connected nodes, the negative edges increase the separability. Let us assume two nodes $i,j$ connected with a positive edge, where the label of node $i$ is $k$. Given the negative likelihood loss $\mathcal{L}_{nll}(Y_i,\widehat{Y}_i)_k=-\log(\widehat{y}_{i,k})$, the gradient of node $i$ is defined as
${\partial \mathcal{L}_{nll}(Y_i,\widehat{Y}_i)_k / \partial \widehat{y}_{i,k}}$. Similarly, the gradient of neighboring node $j$ follows:
\begin{equation}
    \nabla_j\mathcal{L}_{nll}(Y_i,\widehat{Y}_i)_k = {\partial \mathcal{L}_{nll}(Y_i,\widehat{Y}_i)_k \over \partial \widehat{y}_{i,k}} \cdot {\partial \widehat{y}_{i,k} \over \partial h^{(L)}_{\textbf{i},k}} \cdot {\partial h^{(L)}_{\textbf{i},k} \over \partial h^{(L)}_{\textbf{j},k}}
\end{equation}
Since $\partial h^{(L)}_{\textbf{i},k} / \partial h^{(L)}_{\textbf{j},k} > 0$ with positive connection, we can infer that node $j$ will get closer to node $i$ proportional to the $|\eta \partial h^{(L)}_{\textbf{i},k} / \partial h^{(L)}_{\textbf{j},k}|$. Vice versa, the gradient has the opposite sign but the same scale.
\end{manualtheorem}

\subsection{Computational Complexity}
Our model consists of two main components; feature extraction with domain disentanglement and adaptive propagation. The first module can be approximated as $\mathcal{O}((A+B) \cdot N \cdot e)$, where $A$ and $B$ refer to the time of forward passing in feature extraction and domain discriminator. $N$ is the input size and $e$ is the number of training epochs. The cost of the second module is dominated by the cost of the GNN, $\mathcal{O}(|\mathcal{E}|\theta_{GNN})$, which is proportional to the number, $|\mathcal{E}|$, of edges in the considered graphs and the number, $\theta_{GNN}$, of trainable parameters. Since our method further employs edge weight retrieval using the coordinates, the complexity becomes $\mathcal{O}(|\mathcal{E}|\theta_{GNN}+|\mathcal{E}|\theta_{Adapt}$). Thus, the cost of the entire module can be $\mathcal{O}\left((A+B) \cdot N \cdot e+|\mathcal{E}|(\theta_{GNN}+\theta_{Adapt})\right)$.

\begin{table}[t]
\caption{Details of the benchmark datasets}
\label{dataset}
\centering
\footnotesize
\begin{tabular}{c|lll}
\multicolumn{1}{l}{}    & \multicolumn{1}{l}{}       &         &                         \\ 
\Xhline{2\arrayrulewidth}
\multicolumn{1}{c|}{Domain}        & Region                       & \# wetland cells & \# cells  \\ 
\Xhline{2\arrayrulewidth}
\multirow{3}{*}{Source} 
                        & Arizona (AZ, sparse)           & 6.05 K (0.6 \%) & 1.05 M        \\
                        & Washington (WA, moderate)             & 23.2 K (2.2 \%) & 1.05 M        \\
                        & Florida (FL, dense)  & 0.12 M (12 \%) & 1.04 M                   \\
\hline
\multirow{3}{*}{Target} & Texas (TX, sparse)            & 1.21 K (0.1 \%) & 1.12 M        \\
                        & Oregon (OR, moderate)               & 15.6 K (1.5 \%) & 1.05 M         \\
                        & Louisiana (LA, dense)    & 0.09 M (8.3 \%) & 1.05 M        \\
\Xhline{2\arrayrulewidth}
\end{tabular}
\end{table}

\section{Experiments}
In this section, we investigate three key questions that characterize the proposed method and provide a comprehensive analysis:
\begin{itemize}[leftmargin=*]
\item \textbf{RQ1:} Does the proposed model achieve good performance compared to state-of-the-art baselines?
\item \textbf{RQ2:} How much do domain disentanglement and adaptive propagation contribute to the overall performance? Does the shareable knowledge help to solve the sparsity issue?
\item \textbf{RQ3:} Does domain disentanglement effectively distinguish between domain-specific and domain-shareable features?
\end{itemize}

\begin{table}[t]
\small
\centering
\caption{ (RQ1) Accuracy ($\%$) and recall ($\%$) for 3 target regions -- bold with underline indicate best accuracy and recall}
\label{tab:performance}
\centerline{
\begin{adjustbox}{width=\columnwidth}
\begin{tabular}{@{}c|c|c|ccc|ccc|ccc@{}}
\cmidrule[2pt]{1-12}
\multirow{2}{*}{} & \multirow{2}{*}{Method} & \multirow{2}{*}{Metric} &  \multicolumn{3}{c|}{\underline{Texas}} & \multicolumn{3}{c|}{\underline{Oregon}} & \multicolumn{3}{c}{\underline{Louisiana}} \\
& &  & AZ & WA & FL & AZ & WA & FL & AZ & WA & FL \\
\cmidrule[1.5pt]{1-12}
\multirow{8}{*}{\rotatebox{90}{Single-Domain}} & \multirow{1}{*}{MLP} & Acc. &  \multicolumn{3}{c|}{82.0} & \multicolumn{3}{c|}{85.9} & \multicolumn{3}{c}{89.3} \\
&  \multirow{1}{*}{(2 layers)} & Rec. &  \multicolumn{3}{c|}{62.6} & \multicolumn{3}{c|}{94.6} & \multicolumn{3}{c}{99.0} \\ \cline{2-12}
& \multirow{1}{*}{GCN} & Acc. &  \multicolumn{3}{c|}{82.8} & \multicolumn{3}{c|}{91.0} & \multicolumn{3}{c}{91.3}  \\ 
& \multirow{1}{*}{\cite{kipf2016semi}}& Rec. &  \multicolumn{3}{c|}{63.1} & \multicolumn{3}{c|}{92.0} & \multicolumn{3}{c}{99.1}  \\  \cline{2-12}
 & \multirow{1}{*}{GAT} & Acc. &  \multicolumn{3}{c|}{80.7} & \multicolumn{3}{c|}{84.7} & \multicolumn{3}{c}{90.4}  \\
 & \multirow{1}{*}{\cite{velickovic2017graph}} & Rec. &  \multicolumn{3}{c|}{68.3} & \multicolumn{3}{c|}{77.8} & \multicolumn{3}{c}{98.8} \\ \cline{2-12}
 & \multirow{1}{*}{FAGCN } & Acc. &  \multicolumn{3}{c|}{82.5} & \multicolumn{3}{c|}{91.8} & \multicolumn{3}{c}{91.5} \\
 &  \multirow{1}{*}{\cite{bo2021beyond}}& Rec. &  \multicolumn{3}{c|}{62.9} & \multicolumn{3}{c|}{86.7} & \multicolumn{3}{c}{99.5} \\ 
\cmidrule[1.5pt]{1-12}
\multirow{8}{*}{\rotatebox[origin=c]{90}{Cross-Domain}} & \multirow{1}{*}{DAREC} & Acc. & 81.2 & 81.3  & 84.0  & 84.4  & 86.5 & 86.2  & 88.4  & 89.4  & 89.4 \\ 
& \multirow{1}{*}{\cite{yuan2019darec}} & Rec. & 62.0 & 62.3  & 68.9  & 79.3  & 82.1 & 96.2  & 96.6  & 97.3  & 96.8 \\  \cline{2-12}
& \multirow{1}{*}{MMT } & Acc. & 82.0 & 83.0  & 83.1  & 86.2  & 86.6 & 86.0  & 89.1  & 91.3  & 91.3 \\ 
& \multirow{1}{*}{\cite{krishnan2020transfer}}& Rec. & 63.7 & 66.0  & 67.7  & 92.3  & 93.6 & 90.5  & 98.5  & 99.1  & 99.4 \\  \cline{2-12}
& \multirow{1}{*}{SER} & Acc. & 83.5 & 84.0  & 84.3  & 87.2  & 86.4 & 86.8  & 91.0  & 90.1  & 89.0 \\ 
& \multirow{1}{*}{\cite{choi2022based}}& Rec. & 66.6 & 67.2  & 67.5  & 91.2  & 91.4 & 92.8  & 99.3  & 98.3  & 97.1 \\  \cline{2-12}
& \multirow{1}{*}{DH-GAT} & Acc. & 83.4 & 85.1  & 86.5  & 88.8 & 90.7 & 90.2  & 90.6  & 91.5  & 91.4 \\ 
 & \multirow{1}{*}{\cite{xu2023decoupled}}  & Rec. & 70.4 & 74.0  & 74.9  & 95.8  & 92.0 & 97.3  & 99.1  & 99.7  & 99.6 \\ 
\cmidrule[1.5pt]{1-12}
 \multirow{2}{*}{\rotatebox[origin=c]{90}{Ours}}& \multirow{2}{*}{PoTA} & Acc. & 84.1 & 85.4 & \underline{\textbf{87.2}} & 92.1 & \underline{\textbf{93.0}} & 92.6 & 92.2 & 92.3 & \underline{\textbf{92.5}}  \\
 &  & Rec. & 70.8 & 74.2 & \underline{\textbf{77.0}} & 96.0 & 97.8 & \underline{\textbf{98.1}} & 99.5 & 99.6 & \underline{\textbf{99.8}}  \\ \cmidrule[2pt]{1-12}
\end{tabular}
\end{adjustbox}
}
\end{table}

\subsection{Datasets, Baselines, and Setup} \label{datasets}
\textbf{(Datasets)} In Section \ref{data_source}, we introduced the details of three datasets; NLCD, Soil, and HAND. In this section, we consider 6 regions of the US with varying wetland characteristics~Table \ref{dataset}. 

\noindent \textbf{(Baselines)} We consider 8 state-of-the-art methods, both single-domain and cross-domain, listed in Table~\ref{tab:performance}. As a hyperparameter, we set $\lambda=0.2$ in Eq. \ref{overall_loss} for our model. In addition, the learning ratio is set as $1e^{-3}$ with the early stopping of 300 epochs for all baselines.

\noindent \textbf{(Experimental setup)} Training/validation/testing sets are split  10\%/40\%/50\% for all methods. We utilized PyTorch, torch-geometric, and a single GPU (Nvidia Titan Xp) for evaluation. For reproducibility, we uploaded the source codes to our \textit{GitHub}\footnote{https://github.com/ChoiYoonHyuk/Wetland}.

\begin{table}[t]
  \centering
  \caption{(RQ1) Wetland prediction accuracy as a function of the wetland density: sparse regions generally lead to lower prediction accuracy and they greatly benefit from knowledge transfer from  a dense region (such as FL)}
  	\begin{adjustbox}{width=1\columnwidth}
    \begin{tabular}{r|cc|cc|cc|}
\cline{2-7}          & \multicolumn{6}{c|}{\textbf{Region}} \\
\cline{2-7}          & \multicolumn{2}{c|}{\textbf{Sparse}} & \multicolumn{2}{c|}{\textbf{Moderate}} & \multicolumn{2}{c|}{\textbf{Dense}} \\
          & \textbf{TX:0.1\%} & \textbf{AZ:0.6\%} & \textbf{OR:1.5\%} & \textbf{WA:2.2\%} & \textbf{LA:8.3\%} & \textbf{FL:12\%} \\
\cline{2-7}    \textbf{w/o KT} & 83.5\% & 61.4\% & 91.4\% & 81.8\% & 91.9\% & 88.4\% \\
    \textbf{KT from FL} & 87.2\% & 66.6\% & 92.6\% & 82.7\% & 92.5\% & 88.4\% \\
\cline{2-7}    \textbf{Gain from KT} & 4.4\% & 8.5\% & 1.3\% & 1.1\% & 0.7\% & 0.0\% \\
\cline{2-7}
    \end{tabular}%
    \end{adjustbox}
  \label{tab:acc1}%
\end{table}%

\begin{table}[t]
  \centering
  \caption{(RQ1) Accuracy gains due to knowledge transfer: gains are higher when the knowledge is transferred from denser regions (LA and FL) to sparser regions (TX and AZ)}
  \begin{adjustbox}{width=1\columnwidth}
    \begin{tabular}{rr|ccccccc}\cline{3-8}
          &       & \multicolumn{6}{c|}{\textbf{Target}}           &  \\
\cline{3-8}          &       & \textbf{TX:0.1\%} & \textbf{AZ:0.6\%} & \textbf{OR:1.5\%} & \textbf{WA:2.2\%} & \textbf{LA:8.3\%} & \multicolumn{1}{c|}{\textbf{FL:12\%}} & \textbf{AVG} \\ \hline
    \multicolumn{1}{c|}{\multirow{6}[1]{*}
   {\rotatebox{90}{\textbf{Source}}}
    } & \textbf{TX:0.1\%} & \cellcolor[rgb]{ .988,  .988,  1}0.0\% & \cellcolor[rgb]{ .871,  .941,  .902}1.1\% & \cellcolor[rgb]{ .914,  .961,  .937}0.7\% & \cellcolor[rgb]{ .957,  .976,  .973}0.3\% & \cellcolor[rgb]{ .949,  .973,  .965}0.4\% & \multicolumn{1}{c|}{\cellcolor[rgb]{ .937,  .969,  .957}0.5\%} & 0.6\% \\
    \multicolumn{1}{c|}{} & \textbf{AZ:0.6\%} & \cellcolor[rgb]{ .925,  .965,  .945}0.6\% & \cellcolor[rgb]{ .988,  .988,  1}0.0\% & \cellcolor[rgb]{ .906,  .957,  .929}0.8\% & \cellcolor[rgb]{ .988,  .988,  1}0.0\% & \cellcolor[rgb]{ .957,  .976,  .973}0.3\% & \multicolumn{1}{c|}{\cellcolor[rgb]{ .957,  .976,  .973}0.3\%} & 0.4\% \\
    \multicolumn{1}{c|}{} & \textbf{OR:1.5\%} & \cellcolor[rgb]{ .765,  .898,  .808}2.1\% & \cellcolor[rgb]{ .765,  .898,  .808}2.1\% & \cellcolor[rgb]{ .988,  .988,  1}0.0\% & \cellcolor[rgb]{ .925,  .965,  .945}0.6\% & \cellcolor[rgb]{ .957,  .976,  .973}0.3\% & \multicolumn{1}{c|}{\cellcolor[rgb]{ .925,  .965,  .945}0.6\%} & 1.1\% \\
    \multicolumn{1}{c|}{} & \textbf{WA:2.2\%} & \cellcolor[rgb]{ .788,  .906,  .827}1.9\% & \cellcolor[rgb]{ .71,  .878,  .761}2.6\% & \cellcolor[rgb]{ .82,  .922,  .855}1.6\% & \cellcolor[rgb]{ .988,  .988,  1}0.0\% & \cellcolor[rgb]{ .949,  .973,  .965}0.4\% & \multicolumn{1}{c|}{\cellcolor[rgb]{ .937,  .969,  .957}0.5\%} & 1.4\% \\
    \multicolumn{1}{c|}{} & \textbf{LA:8.3\%} & \cellcolor[rgb]{ .388,  .745,  .482}5.6\% & \cellcolor[rgb]{ .659,  .855,  .714}3.1\% & \cellcolor[rgb]{ .894,  .953,  .918}0.9\% & \cellcolor[rgb]{ .863,  .937,  .89}1.2\% & \cellcolor[rgb]{ .988,  .988,  1}0.0\% & \multicolumn{1}{c|}{\cellcolor[rgb]{ .914,  .961,  .937}0.7\%} & 2.3\% \\
    \multicolumn{1}{c|}{} & \textbf{FL:12\%} & \cellcolor[rgb]{ .592,  .831,  .659}3.7\% & \cellcolor[rgb]{ .431,  .765,  .522}5.2\% & \cellcolor[rgb]{ .863,  .937,  .89}1.2\% & \cellcolor[rgb]{ .894,  .953,  .918}0.9\% & \cellcolor[rgb]{ .925,  .965,  .945}0.6\% & \multicolumn{1}{c|}{\cellcolor[rgb]{ .988,  .988,  1}0.0\%} & 2.3\% \\
\cline{1-9}          & \textbf{AVG} & 2.8\% & 2.8\% & 1.0\% & 0.6\% & 0.4\% & \multicolumn{1}{c|}{0.5\%} &  \\
    \end{tabular}%
    \end{adjustbox}
  \label{tab:acc2}%
\end{table}%

\begin{figure}[t]
 \includegraphics[width=0.3\textwidth]{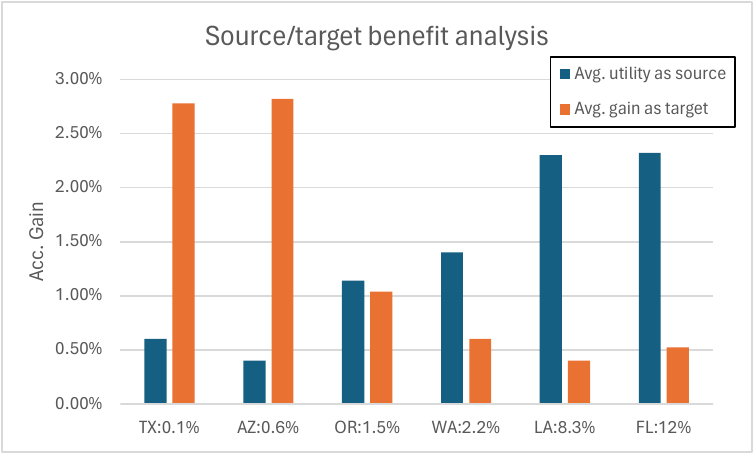}
  \caption{(RQ1) While knowledge transfer always provides gains, sparser regions benefit most from being the target, whereas denser regions are best used as the source domains }
  \label{fig:benefit}
\end{figure}

\subsection{Comparison with Baselines (RQ1)} \label{rq1}
In Table~\ref{tab:performance}, we present accuracy and recall results to compare PoTA to the baselines.
Firstly, we see that PoTA provides the best accuracy and recall. 
We see that, as expected, for the sparse target (Texas), cross-domain methods achieve better results,  while the performance gap is smaller in dense domains (Oregon and Louisiana). In particular, knowledge transfer from the dense (Florida) to the sparse (Texas) domain significantly improves the quality of prediction. 
%
Along with knowledge transfer, we claim that adaptive propagation also plays an important role in performance improvement. This can be observed when using Arizona as the source domain (sparse). Despite our model gaining almost no benefit from knowledge transfer, significant performance improvements can be seen compared to GCN \cite{kipf2016semi} and GAT \cite{velickovic2017graph} across three datasets.

\textbf{(Overview)} 
The high-level overview of the need for knowledge transfer in wetland prioritization can be summarized as follows. Let us first consider Table~\ref{tab:acc1}, where we see that knowledge transfer from a wetland-rich region, FL, is providing significant gains in accuracy, especially in regions, such as TX and AZ, where wetlands are sparse. These results are confirmed in Table~\ref{tab:acc2} and  Figure~\ref{fig:benefit}: while knowledge transfer always provides positive gains in accuracy, sparser regions benefit most from being the target, whereas denser regions are best used as the source.

\begin{table}[!t]
\setlength{\tabcolsep}{1.5em}
\centering
\small
\caption{(RQ2) Accuracy and recall  with ablation for  domain disentanglement (DD) and adaptive propagation (AP)
}

\begin{adjustbox}{width=1\columnwidth}
{
    \begin{tabular}{
        lcccccc
    }
    \toprule
    \toprule
 & \multicolumn{2}{c}{TX (from FL)} & \multicolumn{2}{c}{OR (from WA)} & \multicolumn{2}{c}{LA (from FL)} \\
    \cmidrule(l){2-3} \cmidrule(l){4-5} \cmidrule(l){6-7}
    
Methods    & Acc. & Rec. & Acc. & Rec. & Acc. & Rec. \\
\cmidrule(l){1-2}    \cmidrule(l){2-3} \cmidrule(l){4-5} \cmidrule(l){6-7}
w/o DD & 84.4 & 73.6& 88.3 & 92.1 & 91.4  & 99.4  \\
w/o AP & 84.9 & 76.1 & 87.6  & 90.5  & 90.9  & 99.0 \\
PoTA  & 87.2  & 77.0  & 93.0  & 97.8 & 92.5  & 99.8 \\
    \midrule
    \bottomrule
    \end{tabular}
}
\end{adjustbox}
\label{tab:ablation}
\end{table}

\begin{figure}[t]
 \centerline{
\begin{tabular}{c}
\vspace*{-0.1in}\includegraphics[width=2.5in]{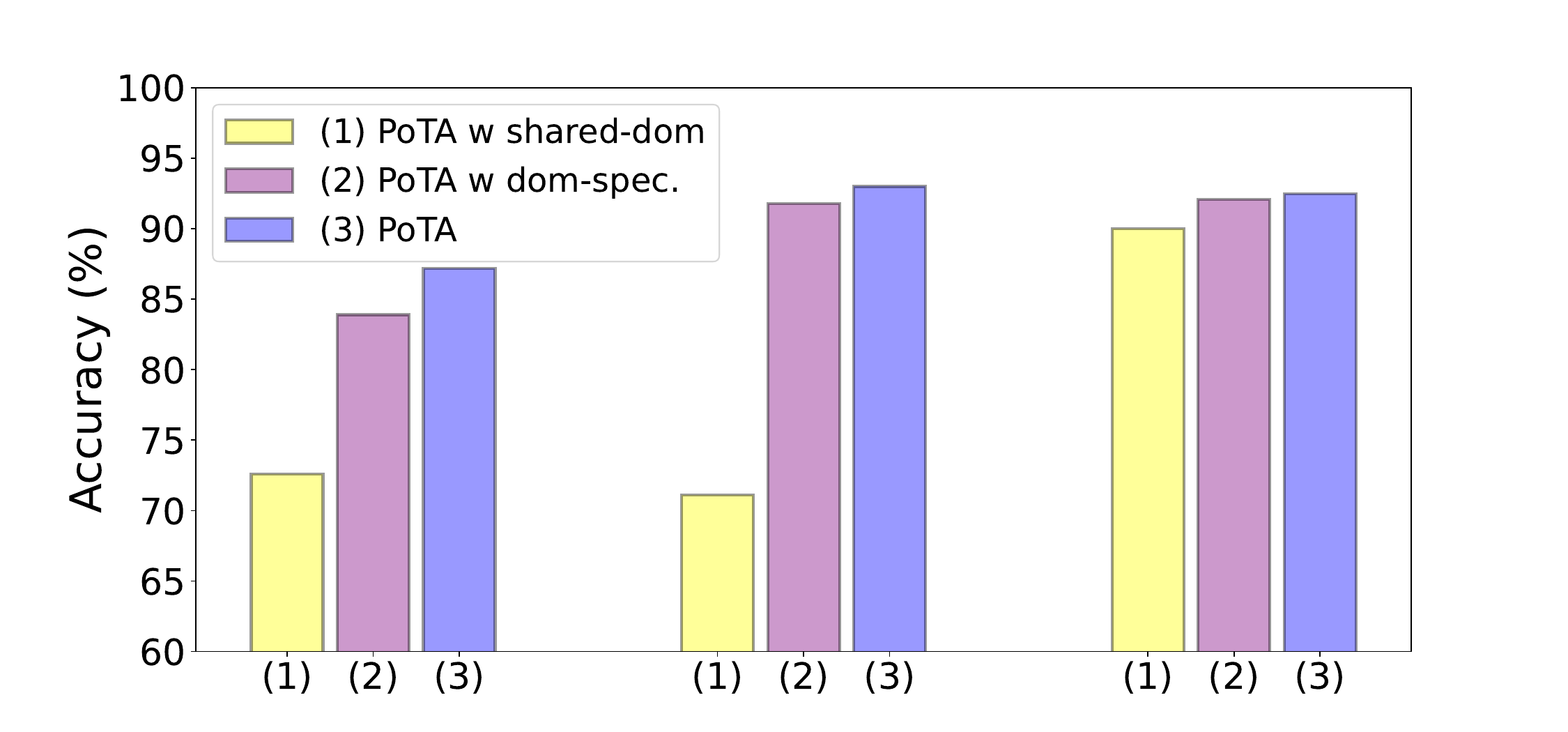}\\
\;\;\;\;\;\;Texas\;\;\;\;\;\;\;\;\;Oregon\;\;\;\;\;Louisiana\\
(a) feature ablation \\
\vspace*{-0.1in}\includegraphics[width=2.5in]{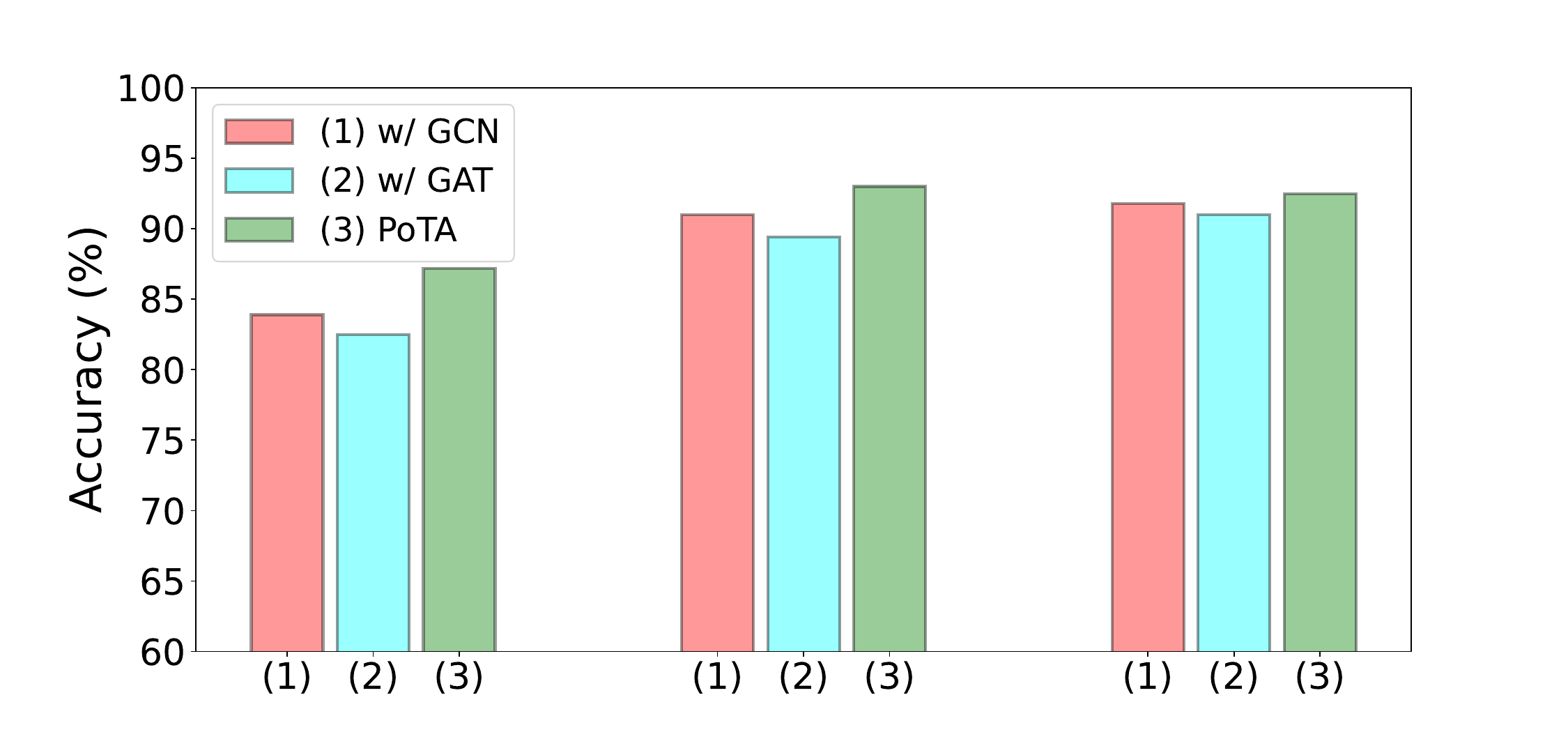}\\
\;\;\;\;\;\;Texas\;\;\;\;\;\;\;\;\;Oregon\;\;\;\;\;Louisiana\\
(b) propagation ablation
\end{tabular} 
 }
  \vspace*{-0.1in}
  \caption{(RQ2)  Accuracy impact of (a)  using domain-specific vs. domain-shareable features and (b)  adaptive propagation}
  \label{fig:case}
\end{figure}

\subsection{Impact of Disentanglement, Sharing, and Adaptive Propagation (RQ2)} \label{rq2}
One exception in Table~\ref{tab:performance} is for the target OR, for which the nearby region, WA, with moderate wetlands, has a greater impact as a source than the wetland-rich region, FL. 
This makes sense as WA is likely to share more with OR, which can help boost predictions in OR even though it has fewer wetlands than FL.
This illustrates the importance of the shared-domain knowledge.

In Table \ref{tab:ablation}, we present an ablation study to test the efficacy of domain disentanglement and adaptive propagation. Specifically, we either remove the domain loss in Eq.~\ref{da_loss} for PoTA (w/o DD) or propagation in Eq. \ref{message_passing} for PoTA (w/o AP). For each of the three targets, we select the source domain that achieved the best performance in Table~\ref{tab:performance}. From the result, we see that excluding domain loss, which hampers region-to-region knowledge transfer, leads to the lowest performance for the sparse domain (Texas). However, it can be found that adaptive propagation plays a more crucial role for denser regions (Oregon and Louisiana) by enabling within-region knowledge transfer.

In Figure~\ref{fig:case}, we further investigate the impacts of the selected latent features (upper) or the propagation scheme (lower). 
As seen in the upper chart, as expected, (2) domain-specific training yields better performance than training only with (1) shared-domain features. However, (3) PoTA effectively combines advantages of the both features for superior performance (Theorem \ref{thm1}).

For the lower chart of Figure~\ref{fig:case}, we replace adaptive propagation with GCN~\cite{kipf2016semi} and GAT~\cite{velickovic2017graph} style propagation: as we see in the chart, adaptive propagation provides the highest accuracy, emphasizing the necessity of distinguishing between the like and unlike neighbors during information exchange (Theorem \ref{thm2}).

\begin{figure}[t]
\centerline{
\begin{tabular}{cc}
\includegraphics[width=1.6in]{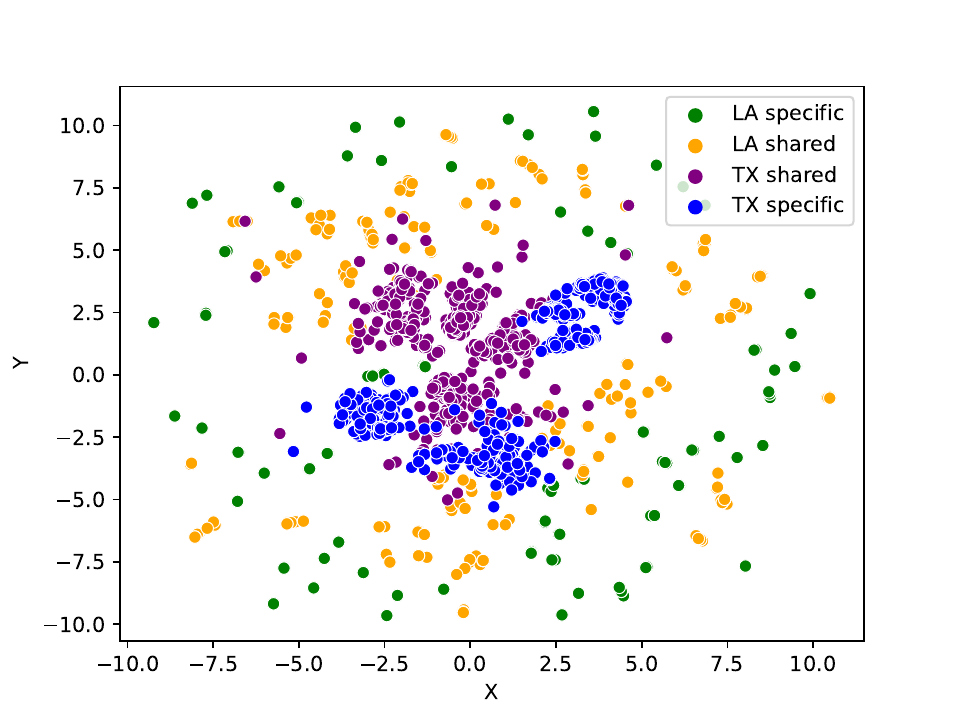}&
\includegraphics[width=1.6in]{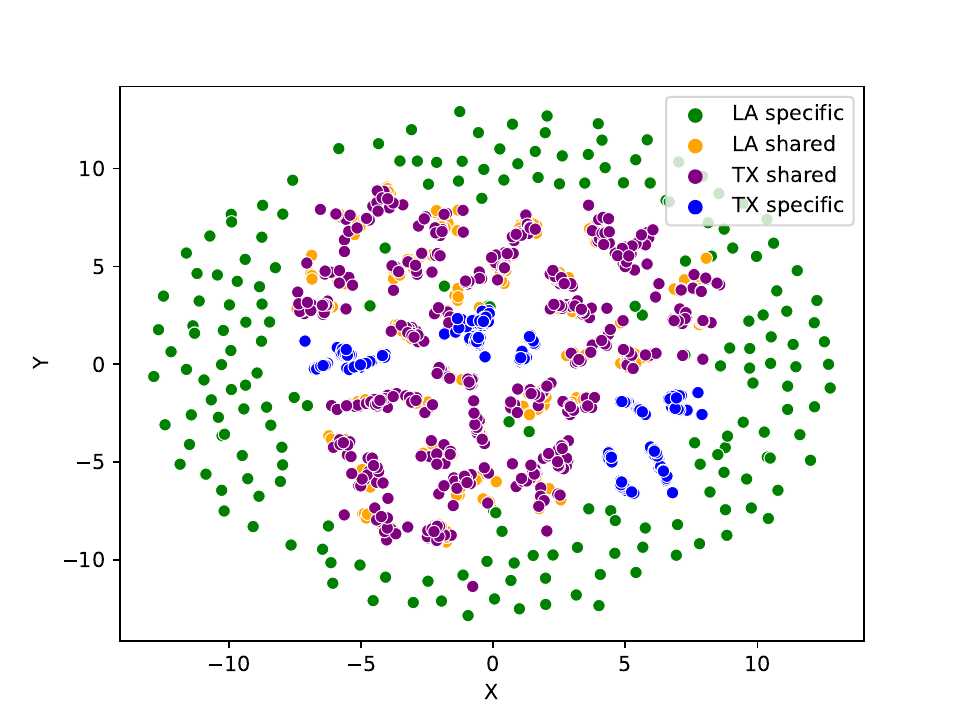}\\
(a) w/o domain disentanglement & (b) w/ domain disentanglement 
\end{tabular}
}
  \vspace*{-0.1in}
  \caption{(RQ3) t-distributed Stochastic Neighbor Embedding (t-SNE) based visualization of the latent features on Texas (TX). Louisiana (LA) is the source domain
}

  \label{fig:viz}
\end{figure}

\subsection{Impact of Feature Disentanglement (RQ3)} \label{rq4}
To demonstrate the contribution of domain disentanglement to the separation of extracted features, we sampled 500 cells from each domain. Then, we visualized the latent vectors from feature extractors in Figure \ref{fig:viz}. Specifically, these vectors comprise source-specific ($l^s_{spe}$), source-shared ($l^s_{shr}$), target-specific ($l^t_{spe}$), and target-shared ($l^t_{shr}$) features. Referring to Figure \ref{fig:viz}a, it is evident that the shared features ($l^s_{shr}$ and $l^t_{shr}$) do not properly overlap. Additionally, some source-specific features ($l^s_{spe}$) overlap with the target-specific ones ($l^t_{spe}$), indicating a lack of effective disentanglement. Conversely, Figure \ref{fig:viz}b demonstrates that shared domain features overlap with each other, while domain-specific features are well separated along with the domain-shared features.

\section{CONCLUSIONS}
Wetlands are essential, yet they are insufficient in many regions. Existing approaches to prioritizing potential future wetland locations rely heavily on expert knowledge. While data-driven techniques show promise to help guide experts, most of the US wetlands account for only about 6\% of the land cover, making data sparsity a major challenge. In this paper, we propose addressing this through region-to-region and within-region knowledge transfer through domain dis-entanglement and adaptive propagation. This led to a 6.1\% improvement in accuracy 
in regions where wetlands are sparse. We believe this could make a significant contribution to dry regions, like the southwestern US.


\bibliographystyle{ACM-Reference-Format}
\bibliography{references.bib}

\end{document}